\documentclass[preprint,prb,preprintnumbers,amsmath,amssymb,showpacs,floatfix]{revtex4}
\usepackage{graphicx,bm,amsmath,epsfig}

\def\bb{\mbox{\protect\boldmath $b$}}

\def\<{\langle}
\def\>{\rangle}

\begin{document}


\def\reff#1{(\ref{#1})}

\def\spose#1{\hbox to 0pt{#1\hss}}
\def\ltapprox{\mathrel{\spose{\lower 3pt\hbox{$\mathchar"218$}}
 \raise 2.0pt\hbox{$\mathchar"13C$}}}
\def\gtapprox{\mathrel{\spose{\lower 3pt\hbox{$\mathchar"218$}}
 \raise 2.0pt\hbox{$\mathchar"13E$}}}

\def\bsigma{\mbox{\protect\boldmath $\sigma$}}
\def\bpi{\mbox{\protect\boldmath $\pi$}}
\def\smfrac#1#2{{\textstyle\frac{#1}{#2}}}
\def\smhalf{ {\smfrac{1}{2}} }

\newcommand{\re}{\mathop{\rm Re}\nolimits}
\newcommand{\im}{\mathop{\rm Im}\nolimits}
\newcommand{\trace}{\mathop{\rm tr}\nolimits}
\newcommand{\fr}{\frac}

\def\Z{{\mathbb Z}}
\def\R{{\mathbb R}}
\def\C{{\mathbb C}}

\title{Polymers as compressible soft spheres}

\author{Giuseppe D'Adamo}
\email{giuseppe.dadamo@aquila.infn.it}
\affiliation{Dipartimento di Fisica, Universit\`a dell'Aquila, 
V. Vetoio 10, Loc. Coppito, I-67100 L'Aquila, Italy}
\author{Andrea Pelissetto}
\email{andrea.pelissetto@roma1.infn.it}
\affiliation{Dipartimento di Fisica, Sapienza Universit\`a di Roma and 
INFN, Sezione di Roma I, P.le Aldo Moro 2, I-00185 Roma, Italy}
\author{Carlo Pierleoni}
\email{carlo.pierleoni@aquila.infn.it}
\affiliation{$^3$ Dipartimento di Fisica, Universit\`a dell'Aquila and 
CNISM, UdR dell'Aquila, V. Vetoio 10, Loc. Coppito, I-67100  L'Aquila, Italy}

\begin{abstract}
We consider a coarse-grained model in which polymers under good-solvent 
conditions are represented by soft
spheres whose radii, which should be identified
with the polymer radii of gyrations, are allowed to fluctuate. 
The corresponding pair potential
depends on the sphere radii. This model is a single-sphere version of the 
one proposed in Vettorel {\em et al.}, Soft Matter {\bf 6}, 2282 (2010),
and it is sufficiently simple to allow us to determine all potentials accurately
from full-monomer simulations of two isolated polymers (zero-density 
potentials). We find that in the dilute regime
(which is the expected validity range of               
single-sphere coarse-grained models based on zero-density potentials)
this model correctly reproduces the density dependence of the radius of 
gyration. However, for the thermodynamics and 
the intermolecular structure, the model is largely equivalent to the 
simpler one in which the sphere radii are fixed to the average 
value of the radius of gyration and radii-independent
potentials are used: for the thermodynamics there is no advantage 
in considering a fluctuating sphere size. 
\end{abstract}



\maketitle

\section{Introduction}

Polymer solutions are very interesting soft-matter systems,
showing a wide variety of behaviors, depending on density, temperature,
architecture, etc.\cite{deGennes,Doi,desCloizeauxJannink,Schaefer}
Due to the large number of atoms belonging to a single macromolecule,
simulations of polymer systems are quite challenging. For this
reason, during the years many attempts have been made to 
develop coarse-grained (CG) models in which only some {\em relevant}
degrees of freedom are retained, in such a way to reproduce some
large-scale structural properties and the thermodynamic behavior.
\cite{Likos:2001p277} 
In the simplest approach one maps polymer chains onto point particles
interacting by means of the pairwise potential of mean force between the 
centers of mass of two {\em isolated} polymers.
\cite{Likos:2001p277,Grosberg:1982p2265,Dautenhahn:1994p2250,Dijkstra:1999p2142}
Since the potentials are computed in the limit of zero polymer density,
this approach is limited to the dilute regime, in which 
many-body interactions \cite{vF-etal-00,Bolhuis:2001p288} can be neglected. 
This limitation was overcome 
ten years ago,\cite{Louis:2000p269,Bolhuis:2001p268} by introducing
pair potentials depending on the polymer density, thus allowing the model
to reproduce exactly the thermodynamics at any given density.
This work has paved the way to the use of soft effective particles to 
represent polymer coils in complex situations such as in modelling 
colloid-polymer mixtures.\cite{Bolhuis:2002p267}
However, deriving density-dependent potentials requires
full-monomer simulations at finite polymer density, which is what one would like
to avoid by using CG models.
Moreover, care is needed to derive the correct thermodynamics 
\cite{HansenMcDonald,Stillinger:2002p2125,Louis:2002p2193} and to compute 
free energies and phase diagrams.\cite{Likos:2001p277} 

These limitations can be overcome by switching to a model at a 
lower level of coarse graining, i.e. by mapping a long linear polymer to a 
short linear chain of soft effective blobs.
\cite{Pierleoni:2007p193,Fritz:2009p1721,Pelissetto:2009p287,%
CG-10,Vettorel:2010p1733,DPP-11,DPP-12} 
Quite recently, extending an older phenomenological
approach,\cite{Murat:1998p1980,EM-01}
Ref.~\onlinecite{Vettorel:2010p1733} proposed a multiblob model. 
As in the models proposed in 
Refs.~\onlinecite{Pierleoni:2007p193,Pelissetto:2009p287,%
DPP-11,DPP-12}, the potentials do not depend on polymer density and are 
fixed by using structural data obtained in the limit of zero polymer density.
However, at variance with the other approaches, the radius of 
each blob, which should be identified with its radius 
of gyration, is not fixed but is allowed to fluctuate.
The idea is very appealing and the model is, in principle, more realistic, 
since blobs are compressible: when the density increases, 
the average blob radius of gyration decreases as it does in the original 
polymer model. Therefore, this model should provide a more 
accurate description of the structural properties of the polymers.
However, its practical implementation is by no means straightforward.
A phenomenological approach was proposed in 
Ref.~\onlinecite{Vettorel:2010p1733}, in which intermolecular 
and intramolecular potentials appropriate for polymers under 
good-solvent conditions were determined by
combining exact results for ideal chains and some approximate predictions
for good-solvent polymers
that allowed the inclusion of the local self-repulsion.

In this paper, we wish to test the approach of 
Ref.~\onlinecite{Vettorel:2010p1733} for polymers under good-solvent conditions.
We consider a single-blob system and carefully compare the results obtained for 
the model with radii-dependent potentials with those obtained for the 
simpler model in which the blob size is fixed.
The simplicity of the CG single-blob model
allows us to compute the pair potentials from full-monomer 
simulations of two isolated chains, avoiding any approximation, thus allowing
us to distinguish between the merits/demerits of the method  from
those of the approximations which are needed to implement it.
Moreover, we can discuss two 
different approaches to the coarse graining: in the first one
each polymer is represented by a soft ``compressible" sphere
located in the polymer center of mass (this is the most common
approach when dealing with linear polymers); 
in the second one the position of the sphere coincides with that 
of the central monomer (star-polymer studies favor this second option).

The paper is organized as follows. 
In Sec.~\ref{sec.2} we define the models we investigate. 
In Sec.~\ref{sec3} we present our results:
in Sec.~\ref{sec.3.1} we discuss the third virial coefficient 
and the zero-density three-body forces, while in Sec.~\ref{sec.3.2}
we compare the results for the compressibility factor,
the center-of-mass (or polymer midpoint) pair distribution function, and 
the distribution of the radius of gyration in the semidilute regime. 
Finally, in Sec.~\ref{sec.4} we present our conclusions.

\section{The models} \label{sec.2}

In this paper polymers are represented as ``compressible" soft spheres
of radius $\sigma$: two spheres of radii $\sigma_1$ and $\sigma_2$,
respectively, interact by means of the pair potential 
$V(\sigma_1,\sigma_2;b)$, where $b$ is the relative distance. 
The radius of each sphere is allowed to fluctuate. We assume that 
the normalized radius distribution for an isolated sphere is given by the 
function $P(\sigma)$.
For a system of $L$ spheres in a 
volume $V$, we consider the partition function
\begin{eqnarray}
&& Z = \int d\sigma_1 P(\sigma_1) \ldots d\sigma_L P(\sigma_L)
     Q_L(\sigma_1,\ldots,\sigma_L) 
\nonumber \\ 
&& Q_L(\sigma_1,\ldots,\sigma_L) = 
   \int_V d^3{\bb}_1\ldots d^3{\bb}_L\,
 \exp\left[-\beta \sum_{i>j} V(\sigma_i,\sigma_j;b_{ij})\right],
\end{eqnarray}
where $\bb_i$ is the position of the $i$-th sphere and 
$b_{ij} = |\bb_i - \bb_j|$. Equivalently, as in 
Ref.~\onlinecite{Vettorel:2010p1733}, we can define a potential 
$\beta V_1(\sigma) = - \ln P(\sigma)$ and write the partition function as
\begin{eqnarray}
&& Z = \int d\sigma_1 d^3{\bb}_1\ldots d\sigma_L  d^3{\bb}_L\,
  \exp\left[-\beta\sum_i V_1(\sigma_i) -
    \beta \sum_{i>j} V(\sigma_i,\sigma_j;b_{ij})\right].
\end{eqnarray}
The distribution $P(\sigma)$ as well as the potentials 
$V(\sigma_1,\sigma_2;b)$ are determined in such a way
to reproduce the thermodynamics and the intermolecular spatial 
distribution of the polymers in the limit of zero polymer density. 
More specifically, we consider a polymer model, in which each
polymer consists of $N$ monomers located in ${\bf r}_1,\ldots {\bf r}_N$.
As usual we define the radius of gyration of the chain as 
\begin{equation}
r^2_g = {1\over 2N^2} \sum_{ij} ({\bf r}_i - {\bf r}_j)^2,
\end{equation}
and its average over all polymer configurations as 
$R^2_g(N) = \langle r^2_g \rangle$. We indicate 
zero-density averages with a hat, so that $\hat{R}_g(N)$ is the 
average radius of gyration of a single isolated polymer. 
It is a function of $N$ and for $N\to \infty$ (scaling limit) it
scales as 
\begin{equation} 
\hat{R}_g(N)= a N^\nu (1 + b N^{-\Delta} + \ldots),
\label{hatRg}
\end{equation} 
where $a$ and $b$ are model-dependent 
constants and $\nu$ and $\Delta$ are universal exponents. For good-solvent 
polymers they are known 
quite precisely:\cite{Clisby-10}
$\nu = 0.587597(7)$, $\Delta = 0.528(12)$.  To define 
$P(\sigma)$ we consider the distribution of the radius of gyration of an 
isolated polymer,
\begin{equation}
  Q(s;N) = \langle \delta (r_g - s) \rangle_1,
\end{equation}
where $\langle \cdot \rangle_1$ is the statistical average over all 
the conformations of an isolated polymer made of $N$ monomers.
In the scaling limit
$N\to\infty$, the adimensional quantity $\hat{R}_g(N) Q(s;N)$
is a universal function of $\sigma = s/\hat{R}_g$,
which we identify with $P(\sigma)$: 
\begin{equation}
\hat{R}_g(N) Q(s;N) = P(\sigma) + O(N^{-\Delta}).
\label{defP}
\end{equation}
Scaling corrections decay as $N^{-\Delta}$, where $\Delta$ is the same 
universal correction-to-scaling exponent appearing in Eq.~(\ref{hatRg}).
The function $P(\sigma)$ satisfies
\begin{equation}
\int d\sigma\, P(\sigma) = 1, \qquad\qquad
\int d\sigma\, \sigma^2 P(\sigma) = 1,
\end{equation}
the second equation being a direct consequence of the definition of 
$\hat{R}_g$. To define the potential, we first consider the model in which 
the CG spheres are 
located in the centers of mass of the polymers. In this case
we first consider
\begin{equation}
\beta v(s_1,s_2;r;N) = 
   - \ln {\langle e^{-\beta U_{\rm interm}} 
            \delta (r_{g,1} - s_1) \delta (r_{g,2} - s_2) 
           \rangle_{{\bf 0},{\bf r}} \over 
          Q(s_1;N) Q(s_2;N)},
\end{equation}
where the average
$\langle\cdot \rangle_{{\bf 0},{\bf r}}$ is over all isolated polymer pairs
made of $N$ monomers, 
such that their centers of mass are in the origin ${\bf 0}$ and in ${\bf r}$,
respectively, $r_{g,1}$ and $r_{g,2}$ are the gyration radii of the
two polymers, and $U_{\rm interm}$ is the intermolecular interaction
energy. In the scaling limit $N\to\infty$, $v(s_1,s_2;r;N)$ converges to a 
universal function, which we identify with the pair potential of the 
CG model: 
\begin{equation}
v(s_1,s_2;r;N) = 
V(\sigma_1,\sigma_2;b)  + O(N^{-\Delta}), 
\label{V-def}
\end{equation}
where $\sigma_1 = s_1/\hat{R}_g$, $\sigma_2 = s_2/\hat{R}_g$, and
$b = r/\hat{R}_g$.
Because of these definitions,
the polymer center-of-mass pair distribution function in the limit of zero 
polymer density is the same as the zero-density pair distribution function 
between the centers of the spheres.
Hence, the polymer thermodynamics is correctly reproduced in the low-density 
limit by the effective model.

\begin{widetext}
\begin{figure}
\begin{center}
\begin{tabular}{cc}
\epsfig{file=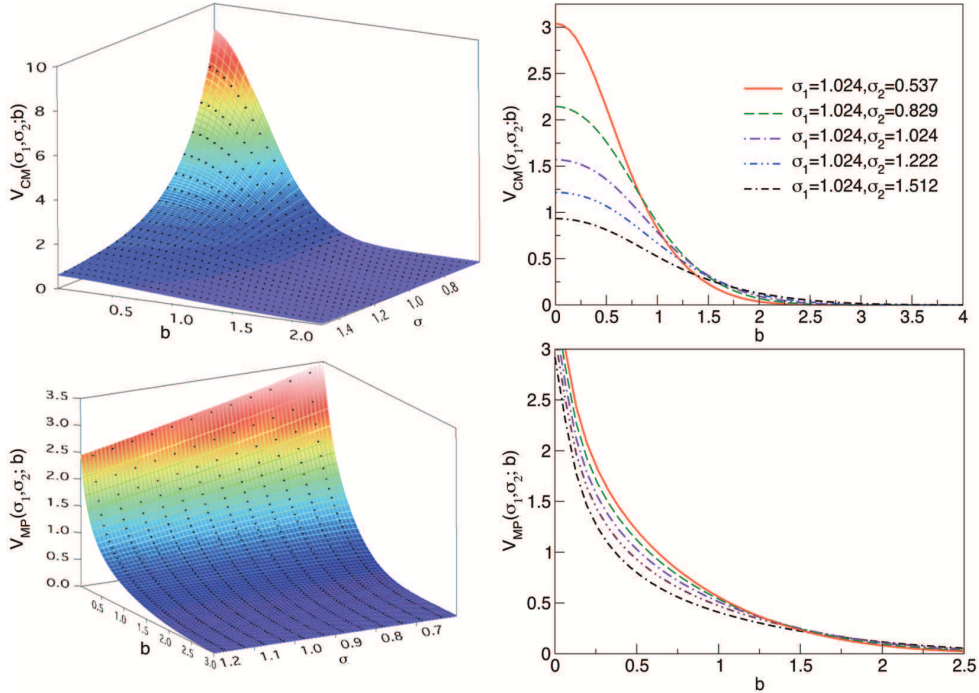,angle=0,width=13truecm} \hspace{0.5truecm} 
\end{tabular}
\end{center}
\caption{On top we report the center-of-mass 
potentials $\beta V(\sigma_1,\sigma_2;b)$: on the left we show a 
three-dimensional plot in terms of $\sigma = \sigma_1=\sigma_2$ and 
$b=r/\hat{R}_g$, on the right we show the potentials 
for $\sigma_1 = 1.024$ and 
several values of $\sigma_2$, as a function of $b$.
On bottom we show the same plots for the midpoint potentials 
$V_{MP}(\sigma_1,\sigma_2;b)$.
}
\label{fig:potCM}
\end{figure}
\end{widetext}

It is important to stress that the quantities $P(\sigma)$ and 
$V(\sigma_1,\sigma_2;b)$ are {\em universal}, i.e. the same result is obtained by using
any model that is appropriate to describe polymers under 
good-solvent conditions. One could use the well-known lattice 
self-avoiding walk model or any off-lattice model appropriate to describe 
good-solvent polymers, for instance the bead-rod model of 
Ref.~\onlinecite{Vettorel:2010p1733}. We will use the lattice 
Domb-Joyce model with $w =0.505838$ 
(see Ref.~\onlinecite{Caracciolo:2006p587} for details on the model).
In the scaling limit $N\to \infty$ such a model describes 
polymers under good-solvent conditions. Morever, for our particular choice 
of the parameter $w$, the scaling corrections proportional to 
$N^{-\Delta}$ that appear in Eqs.~(\ref{defP}) and (\ref{V-def}) are very small,
so that we can obtain the asymptotic (scaling) functions 
$P(\sigma)$ and $V(\sigma_1,\sigma_2;b)$ from simulations of chains of 
moderate length. The results we shall present are obtained by using 
chains with $N=600$. We have also performed simulations with $N=2400$: 
the corresponding results are fully compatible with those 
obtained using $N=600$, indicating the absence of relevant finite-length
effects.  In practice, we simulate two independent Domb-Joyce chains 
using the highly-efficient pivot algorithm
\cite{pivot} (we perform $1.25\times 10^9$ pivot trial moves on each of them)
and determine numerically $Q(s,N=600)$ and 
$v(s_1,s_2;r;N=600)$. Then, we define
$P(\sigma) = \hat{R}_g(N=600) Q(s,N=600)$ and
$V(\sigma_1,\sigma_2;b) = v(s_1,s_2;r;N=600)$, with 
$s = \sigma \hat{R}_g(N=600)$ (analogous relations hold for $s_1$ and $s_2$)
and $r = b \hat{R}_g(N=600)$.
Plots of the 
potential for several values of 
$\sigma_1$ and $\sigma_2$ are reported in Fig.~\ref{fig:potCM}. They have an 
approximately Gaussian behavior with $V(\sigma_1,\sigma_2;b=0)$ increasing
as the radii decrease. This is of course expected, since the smaller
$\sigma_1$ and $\sigma_2$ are, the more compact the two walks become. 
Hence repulsion should increase. For the same reasons, the range of the 
potentials decreases as the radii decrease. 

\begin{figure}
\begin{center}
\begin{tabular}{c}
\epsfig{file=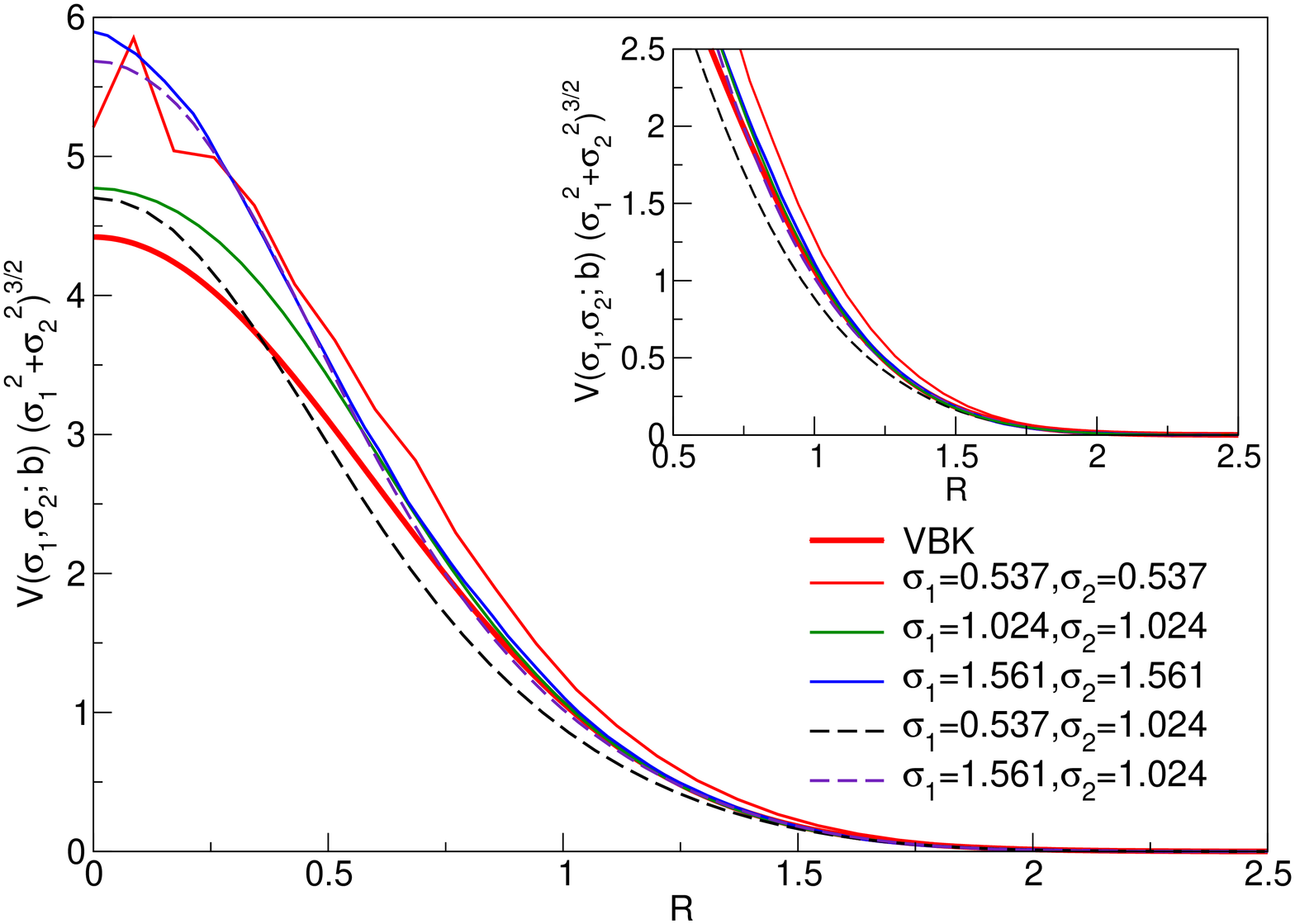,angle=0,width=8truecm} \hspace{0.5truecm} \\
\end{tabular}
\end{center}
\caption{Rescaled potentials 
$(\sigma_1^2 + \sigma_2^2)^{3/2} V(\sigma_1,\sigma_2;b)$ 
as a function of $R = b (\sigma_1^2 + \sigma_2^2)^{-1/2}$
for several values of $\sigma_1$ and $\sigma_2$.
We also plot the function $\epsilon\, e^{-\alpha R^2}$, with 
$\epsilon = 4.42$, $\alpha = 1.42$ (VBK).
}
\label{fig:Vettorel}
\end{figure}

In Ref.~\onlinecite{Vettorel:2010p1733}, 
on the basis of a heuristic argument,
it was suggested that the pair potentials could be approximately written as 
\begin{equation}
V_{VBK}(\sigma_1,\sigma_2;b) = 
  \epsilon (\sigma^2_1 + \sigma_2^2)^{-3/2} 
  \exp\left(-\alpha {b^2\over \sigma^2_1 + \sigma_2^2}\right),
\label{VVBK}
\end{equation}
where $\alpha = 3/2$ and $\epsilon$ is a constant independent of 
$\sigma_1$ and $\sigma_2$. We find that
this expression works reasonably well when taking $\epsilon \approx 4$-5. 
To obtain an optimal approximation we determine $\alpha$ and $\epsilon$ 
in such a way to minimize the functional 
\begin{equation}
\Psi(\alpha,\epsilon) = \int_0^\infty d\sigma_1 \int_0^\infty d\sigma_2 
  \int_0^{b_{\rm max}} db
  \left[P(\sigma_1) P(\sigma_2) b^2 
        \left(e^{-\beta V(\sigma_1,\sigma_2;b)} - 
              e^{-\beta V_{VBK}(\sigma_1,\sigma_2;b)} \right)\right]^2,
\label{Psi}
\end{equation}
where $V(\sigma_1,\sigma_2;b)$ and $P(\sigma)$ are the quantities
computed from full-monomer simulations. 
Since the Monte Carlo estimates of $V$ for $b\gtrsim 3$ are noisy,
we excluded these values from the $b$ integration, taking $b_{\rm max} = 3$.
The functional (\ref{Psi}) 
has been chosen on the basis of the expression of the 
second-virial universal combination\cite{footnotevir}
\begin{equation}
A_2 = 2\pi \int d\sigma_1 d\sigma_2 db 
   P(\sigma_1) P(\sigma_2) b^2 
   \left(1 - e^{-\beta V(\sigma_1,\sigma_2;b)}\right).
\end{equation}
We obtain
\begin{equation}
\alpha = 1.42, \qquad \qquad \epsilon = 4.42.
\label{VVBK-par}
\end{equation}
Note that the optimal value for $\alpha$ is quite close to the value 
$\alpha = 3/2$
proposed in Ref.~\onlinecite{Vettorel:2010p1733}. 
In Fig.~\ref{fig:Vettorel} we plot
$(\sigma_1^2 + \sigma_2^2)^{3/2} V(\sigma_1,\sigma_2;b)$ vs
$R = b (\sigma_1^2 + \sigma_2^2)^{-1/2}$ for several values of $\sigma_1$
and $\sigma_2$. The results are then compared 
with the phenomenological expression 
$\epsilon\, e^{-\alpha R^2}$, obtained using 
potential (\ref{VVBK}). We observe reasonable agreement for 
$R\gtrsim 1$, while significant discrepancies are observed for $R\to 0$. 
However, this is exactly the region which does not contribute significantly
to $A_2$, hence to the thermodynamics. 

Up to now we have characterized the polymer position by using its center of 
mass: in definition (\ref{V-def}) $b$ represents the 
distance between the two centers of mass, expressed in units of $\hat{R}_g$:
$b = r/\hat{R}_g$.
This is the usual choice for CG models of linear polymers.
On the other hand, when considering star polymers, it is much more common
to consider CG models in which the star position 
is identified with the position of the ramification point, the center of the 
star.\cite{LLWAJAR-98,WLL-99,JDLFL-01,DLL-02,Likos:2001p277}
If we view linear polymers as two-arm star polymers, it is 
natural to define the CG model using the central monomer 
as polymer position. 
In an exact mapping this choice would be ininfluential for the 
thermodynamics.\cite{Likos:2001p277}
However, since 
all $n$-polymer interactions with $n\ge 3$ are neglected in the 
CG model, different choices of the point associated
with the polymer position give rise to CG models 
with different thermodynamic behavior.
It makes therefore sense to study the different possibilities, 
with the purpose of understanding which is the optimal one. 

We thus also study two different CG models in which the 
polymer position is given by the central monomer ${\bf r}_{N/2}$. 
We consider a model
in which polymers are represented by identical spheres
interacting by means of the potential of mean force
\begin{equation}
V_{MP}(b) = - \ln 
\langle e^{-\beta U_{\rm interm}}
           \rangle_{{\bf 0},{\bf r}}, 
\label{Vc-def}
\end{equation}
where now the average is over all isolated polymer pairs such that the central 
monomers
(midpoints, MP)
are in the origin and in $\bf r$, respectively,
and $b = r/\hat{R}_g$. We also consider the 
model with compressible soft polymers interacting by means of the  potential 
$V_{MP}(\sigma_1,\sigma_2;b)$ defined as in Eq.~(\ref{V-def}), where now
$b$ is the distance between the central monomers expressed in units of 
$\hat{R}_g$. 

The potential $V_{MP}(b)$ has been discussed at length in the context of 
star polymers. For $b\to 0$ it diverges logarithmically as\cite{WP-86}
$\ln (1/b)$. An explicit parametrization has been given in 
Ref.~\onlinecite{HG-04} (see their results for a two-arm star polymer):
\begin{equation}
V_{MP}(b) = {1\over \tau} 
\ln \left[ \left({\alpha\over b}\right)^{\tau \beta} e^{-\delta b^2} + 
    \exp(\tau \gamma  e^{-\delta b^2})\right],
\label{GH-parametrization}
\end{equation}
where 
\begin{equation}
\alpha = 1.869,\,\, \beta = 0.815,\,\, \gamma=0.372,\,\, \delta = 0.405,\,\, 
\tau = 4.5. 
\label{Vc-parameters}
\end{equation}
This parametrization is quite precise. For instance, we obtain 5.51 for the 
second-virial combination $A_2$, which is very close to the polymer 
result\cite{Caracciolo:2006p587} $A_2 = 5.500(3)$. 

In the bottom panels of Fig.~\ref{fig:potCM} we report the potential 
$V_{MP}(\sigma_1,\sigma_2;b)$. It is interesting to observe
that they all diverge logarithmically as $b\to 0$, apparently
with the same type of logarithmic behavior, 
$V_{MP}(\sigma_1,\sigma_2;b) \sim 0.82 \ln (1/b)$, 
for all values of $\sigma_1$ and $\sigma_2$.

\section{Comparison of the models}  \label{sec3}

In this section we compare the thermodynamic behavior of the models we have 
introduced in the previous section:
\begin{itemize}
\item[a)] the standard CG model in which polymers
are identical soft spheres interacting with a potential which only
depends on distance. We consider the case in which the polymer position
is given by the position of the center of mass
(model M1a) or of the central monomer (model M1b). In the first case 
we use the accurate expression
of the pair potential\cite{footnotepot}
given in Ref.~\onlinecite{Pelissetto:2005p296}, in the second one 
we use Eq.~(\ref{GH-parametrization}) with parameters (\ref{Vc-parameters}). 
\item[b)] 
We consider the ``compressible" CG model in which we use either 
the center of mass (model M2a) or the central monomer (model M2b) 
as polymer position. We also consider the model with pair potential
(\ref{VVBK}) with parameters given by (\ref{VVBK-par}) 
(model M2c).
\end{itemize}
The results will be compared with full-monomer results and, for comparison,
with those obtained in the tetramer model (results will be labelled with 
``t") introduced recently in Ref.~\onlinecite{DPP-11}.

\subsection{Three-body interactions at zero density} \label{sec.3.1}

\begin{table}
\caption{Virial-coefficient 
universal combinations for the models introduced in Sec.~\ref{sec3} and
for the tetramer model (t) of Ref.~\protect\onlinecite{DPP-11}. We also
report the universal asymptotic values for 
polymers (p).\protect\cite{Caracciolo:2006p587}}
\label{A2A3-table}
\begin{tabular}{cccccccc}
\hline\hline
 &  p   & M1a    &   M1b     &            M2a   &  M2b  &  M2c   &    t  \\
\hline
$A_2$ &  5.500(3)& 5.4926(1)&5.5109(1)&5.5102(3) &5.5085(2)&5.5738(3)&5.597(1)\\
$A_3$ &  9.80(2) &  7.844(6)& 4.925(4)&7.42(2)   &4.43(2)  &7.22(2)  &9.99(2)\\
$A_3'$ &10.64    &  7.844(6)& 4.925(4)&8.015(5)  &5.012(2) &7.809(5) &10.57(2)\\
$A_{3,fl}$ & $-$0.84&  0    &  0 &$-$0.59(2)&$-$0.58(2)&$-$0.58(2)&$-$0.581(5)\\
\hline\hline
\end{tabular}
\end{table}

If we expand the compressibility
factor in powers of the concentration $c=L/V$ as
\begin{equation}
Z = {\Pi\over k_B T c} = 1 + B_2 c + B_3 c^2 + O(c^3),
\end{equation}
the quantity $A_2 = B_2/\hat{R}_g^3$ is universal. An accurate estimate
is \cite{Caracciolo:2006p587} $A_2 = 5.500(3)$. 
Since we have matched the center-of-mass or polymer-midpoint 
distribution function to
determine the pair potential, all models should
give the correct estimate of the combination $A_2$.
In Table \ref{A2A3-table} we report the results for the models we consider.
Differences are small, 
and are representative of the level of precision with which the
models reproduce the polymer center-of-mass or midpoint
distribution function. Note that the estimate corresponding to model
M2c is very close to the correct one, indicating that expression
(\ref{VVBK}) parametrizes quite accurately the $R_g$ dependence of the 
potentials.

Much more interesting is the comparison of the third virial coefficient,
since it provides an indication of the accuracy with which the
CG models reproduce the polymer thermodynamics in the dilute regime
and also of the importance of the three-body forces which have
been neglected. The universal combination $A_3 = B_3/\hat{R}_g^6$
was computed in Ref.~\onlinecite{Caracciolo:2006p587} finding
\begin{equation}
A_3 = 9.80(2).
\end{equation}
In order to determine $A_3$, two contributions had to be computed. One
contribution is the standard one, which is the only one present in monoatomic
fluids and in fluids of rigid molecules, $A'_3 \approx 10.64$,
while the second one is
a flexibility contribution $A_{3,fl} \approx -0.84$ (it corresponds to
$-T_1 \hat{R}_g^{-6}$ in the notations of 
Ref.~\onlinecite{Caracciolo:2006p587}).
The combination $A_3$ as well as the two contributions
$A'_3$ and $A_{3,fl}$ are universal, hence it makes sense to compare them with
the corresponding results in the CG models.

We have estimated $A_3$ for all models. 
The results are reported in Table~\ref{A2A3-table}.
Note that, while $A_{3,fl}$ vanishes
in models M1a and M1b, a nonvanishing contribution
with the correct sign is obtained for the models with a fluctuating radius.
On the other hand, the estimates of $A_3$ for 
models with radii-dependent potentials are essentially equivalent to those in
which the sphere radii are fixed:
the estimates corresponding  to models M2a and M2b 
are close to those of models M1a and M1b, respectively. 
To be precise, discrepancies increase by allowing the radii to fluctuate:
the difference between the 
M2a (or M2b) estimate of $A_3$ and the asymptotic polymer result is larger 
than the discrepancy observed for model M1a (or M1b, respectively).
We can also compare the results for models M2a and M2c: the difference 
between the two estimates of $A_3$ is small, confirming that  
expression (\ref{VVBK}) is reasonably accurate.
Finally, the results show that the CG models in which the center of mass is 
taken as reference point are more accurate than those 
in which the central monomer is considered. In the latter case,
$A_3$ is underestimated by approximately a factor of two.

\begin{figure}
\begin{center}
\begin{tabular}{c}
\epsfig{file=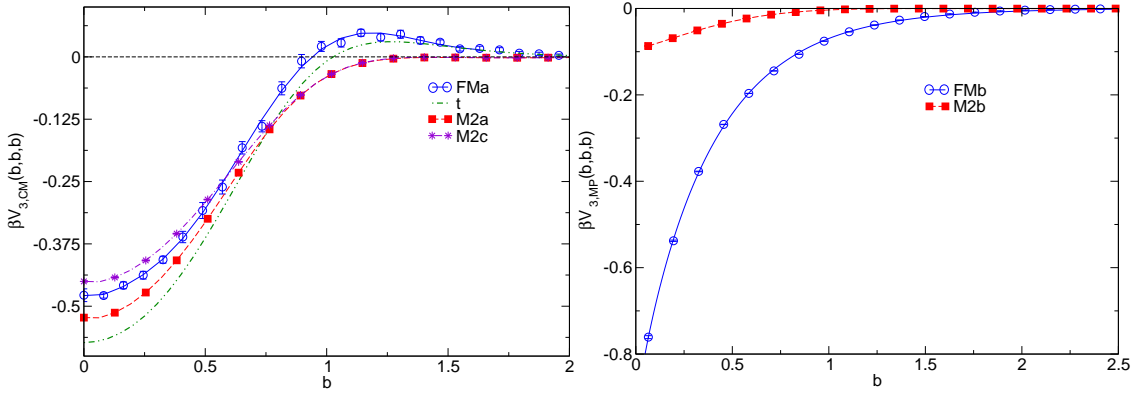,angle=-0,width=15truecm} \hspace{0.5truecm} \\
\end{tabular}
\end{center}
\caption{Three-body potential of mean force
$\beta V_3({\bf r}_{12},{\bf r}_{13},{\bf r}_{23})$ for
${r}_{12}={r}_{13} = {r}_{23} = r$, as a function of
$b = r/\hat{R}_g$. On the left we report results for models M2a, M2c,
for the tetramer model (t) of Ref.~\protect\onlinecite{DPP-11},
and the predictions of full-monomer simulations for the quantity
associated with the center of mass (FMa);
on the right we report the results for model M2b and the 
prediction of full-monomer simulations for the analogous quantity
associated with the polymer midpoint (FMb).
}
\label{threebody}
\end{figure}

As a further check we compute the effective three-body potential of
mean force defined by \cite{Bolhuis:2001p288,Pelissetto:2005p296}
\begin{equation}
\beta V_3({\bb}_{12},{\bb}_{13},{\bb}_{23}) =
- \ln {\langle e^{-\beta U_{12} - \beta U_{13} - \beta U_{23}}
    \rangle_{{\bf b}_{12},{\bf b}_{13},{\bf b}_{23}} \over
    \langle e^{-\beta U_{12}}\rangle_{{\bf b}_{12}}
    \langle e^{-\beta U_{13}}\rangle_{{\bf b}_{13}}
    \langle e^{-\beta U_{23}}\rangle_{{\bf b}_{23}}  };
\label{def-V3}
\end{equation}
here $U_{ij}$ is the intermolecular potential energy
between molecules $i$ and $j$
and the average
$\langle \cdot \rangle_{{\bf b}_{12},{\bf b}_{13},{\bf b}_{23}}$
is over the radii distributions of triplets of isolated  spheres such that
${\bf b}_{ij} = {\bf b}_{i} - {\bb}_{j}$, where
${\bb}_{i}$ is the position of sphere $i$. 
In models M1a and M1b, the sphere radius does not fluctuate, 
so that $\langle e^{-\beta U_{ij}}\rangle_{{\bf b}_{ij}} = 
e^{-\beta V(b_{ij})}$ and 
$\langle e^{-\beta U_{12} - \beta U_{13} - \beta U_{23}}
    \rangle_{{\bf b}_{12},{\bf b}_{13},{\bf b}_{23}} = 
   e^{-\beta V(b_{12}) - \beta V(b_{13}) - \beta V(b_{13}) }$.
It follows that 
$V_3({\bb}_{12},{\bb}_{13},{\bb}_{23}) = 0$.
For models M2a, M2b, and M2c the average is over the distribution $P(\sigma)$
and $U_{ij}$ should be identified 
with $V(\sigma_i,\sigma_j;b_{ij})$.
Potential (\ref{def-V3})
should be compared with the analogous polymer quantity 
in which $\langle \cdot \rangle_{{\bf b}_{12},{\bf b}_{13},{\bf b}_{23}}$ 
is the average over all conformations of triplets of isolated polymers such that 
${\bb}_{i}$ is the position (in units of $\hat{R}_g$) 
of the center of mass of polymer $i$ 
(this is the case relevant for models M1a, M2a, and M2c) or 
the position of the polymer midpoint
(the relevant potential for models M1b and M2b). The polymer three-body potential of mean
force is universal in the scaling limit, i.e. it is model independent. 

We computed $\beta V_3({\bb}_{12},{\bb}_{13},{\bb}_{23})$
for equilateral triangular configurations
such that ${b}_{12}={b}_{13} = {b}_{23} = b$ 
for models M2a and M2b. 
The results are reported
in Fig.~\ref{threebody} and compared with the corresponding quantities
(FMa and FMb) computed in full-monomer simulations.
They were obtained by considering triplets of Domb-Joyce walks made of 
$N=600$ beads. We used the pivot algorithm and performed $2.5\times 10^8$ pivot
trial moves on each of them. We also performed simulations with $N=2400$,
verifying the absence of finite-length effects.

\begin{figure}
\begin{center}
\begin{tabular}{c}
\epsfig{file=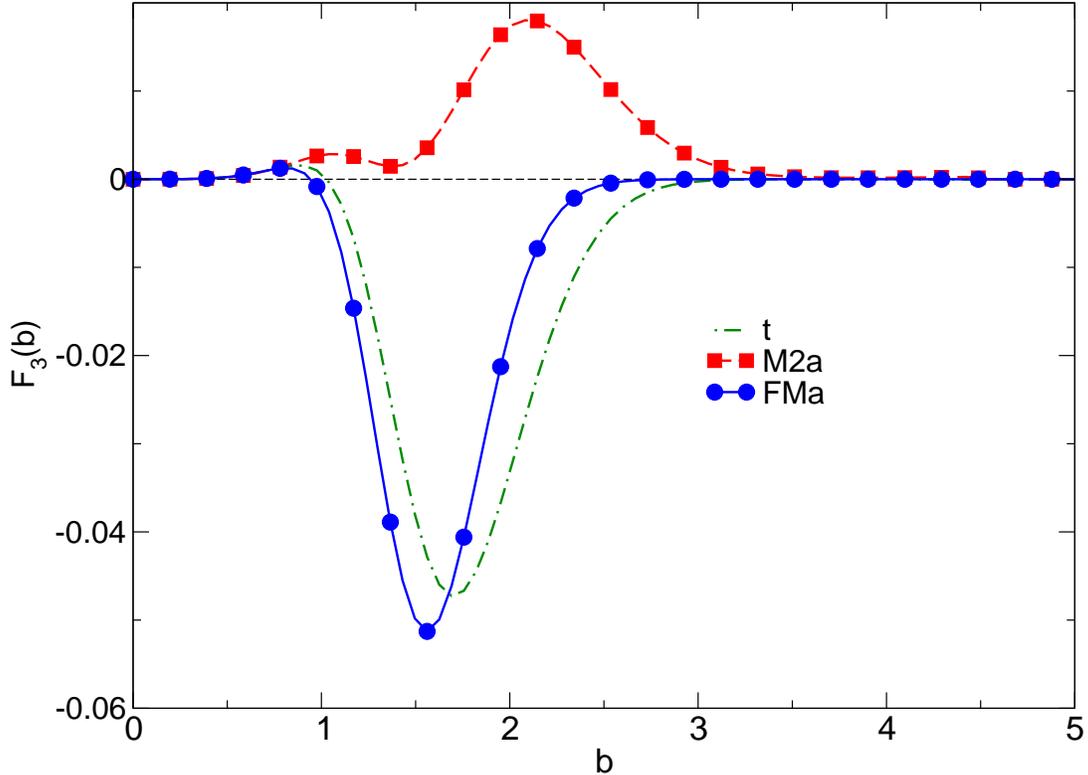,angle=-0,width=15truecm} \hspace{0.5truecm} \\
\end{tabular}
\end{center}
\caption{Function $F_3(b)$ as a function of $b$, for 
polymers (FMa), for model M2a, and for the 
tetramer model of Ref.~\protect\onlinecite{DPP-11} (t).
}
\label{F3}
\end{figure}

For model M2a, we find $\beta V_3(b,b,b) = 0$ for $b \gtrsim 1.2$:
for these values of $b$, model M2a is not different from the 
simpler model M1a. In particular, it does not reproduce the 
repulsive maximum that occurs for $b \approx 1.2$. On the other hand,
model M2a appears to reproduce quite well the attractive short-distance
part of $\beta V_3(b,b,b)$. 
The fact that model M2a gives a better estimate
of $\beta V_3(b,b,b)$ than model M1a, while, at the same time, 
providing
a slightly less accurate estimate of $A_3$ may seem contradictory
at first sight. To understand it, let us note that (see Appendix 
for the derivation) 
\begin{eqnarray}
A_{3,\rm pol} - A_{3,CG}  &=& - {1\over 3} \int d^3{\bb}_{12} d^3{\bb}_{13}
\left(e^{-\beta V_{3,\rm pol}({\bf b}_{12},{\bf b}_{13},{\bf b}_{23})} - 
  e^{-\beta V_{3,\rm CG}({\bf b}_{12},{\bf b}_{13},{\bf b}_{23})} \right)
   \nonumber \\
&& \times 
   g(b_{12}) g(b_{13}) g(b_{23}),
\label{B3diff}
\end{eqnarray}
where  the subscripts ``pol" and "CG" refer to the polymer and the 
CG model, respectively, and $g(b)$ is the zero-density 
center-of-mass
pair distribution function, which is, by definition, identical in the 
polymer and in the CG model. Since $g(b)$ is small for small
values of $b$, keeping also into account that $d{\bb}_{12} d{\bb}_{13}$ 
gives a factor $b_{12}^2 b_{13}^2$, 
the small-distance behavior of 
$V_3(\bb_{12},\bb_{13},\bb_{23})$ is irrelevant for the 
computation of $A_3$. Hence it is much more interesting to compare
\begin{eqnarray}
F_3(b) = b^4 (e^{-\beta V_3(b,b,b)} - 1) g^3(b).
\label{defF3}
\end{eqnarray}
Such a quantity is reported in Fig.~\ref{F3} and shows that the 
relevant region corresponds to $1\lesssim b \lesssim 3$. Moreover,
while $F_3(b)$ is mostly negative for polymers, we have $F_3(b) = 0$ for
model M1a, and, even worse, $F_3(b) > 0$ for model M2a. Hence, 
the compressible model gives a correction to the results for model M1a
which has the wrong sign: hence, the discrepancy with the polymer 
results increases. For comparison, we also include the estimate of 
$F_3(b)$ for the
tetramer model of Ref.~\protect\onlinecite{DPP-11}, which is quite close 
to the polymer result, confirming that the tetramer model is a very good 
CG model in the dilute regime.

In Fig.~\ref{threebody} we also consider model M2b and the 
corresponding full monomer quantity (FMb). Potential FMb
shows a clear logarithmic divergence as $b \to 0$, and indeed
a general theoretical argument \cite{vF-etal-00,P-11} 
predicts $V_3(b,b,b) \simeq - 0.248 \ln (1/b)$ for $b \to 0$.
A fit of the small-distance data to $V_3 \simeq - 0.248 \ln (a/b)$ gives 
$a = 1.17$. The corresponding expression gives a good 
fit of the full-monomer data for $V_3$ up to $b \approx 0.9$. 
The three-body potential $V_3$  for model M2b is significantly 
different from the polymer one. For instance it is finite for 
$b\to 0$: $V_3(0,0,0) \approx -0.1$ for model M2b. Clearly,
the CG model M2b is unable to correctly reproduce the 
three-body interactions.

\subsection{The semidilute regime} \label{sec.3.2}

We now analyze the behavior of the models in the semidilute regime. 
For this purpose we have performed simulations of the CG models for several 
values of the polymer volume fraction 
\begin{equation}
\Phi  = {4\pi\over 3} \hat{R}_g^3 c,
\end{equation}
where $c = L/V$ is the number of polymers per unit volume.
We have determined the compressibility factor $Z = \beta P/c$, the 
inverse compressibility $K = \partial (cZ)/ \partial c$, the 
finite-density adimensional distribution $P(\sigma,\Phi)$ of $r_g$
($\sigma = r_g/\hat{R}_g$ as before), and the ratio 
$S_g(\Phi) = R_g^2(\Phi)/\hat{R}_g^2$.
The compressibility factor was computed by 
using the molecular virial route \cite{Ciccotti:1986p2263,Akkermans:2004p2261}
and checked by comparing it with the (significantly less precise)
result obtained by using the compressibility route (we compute $K$ as 
described in Ref.~\onlinecite{Pelissetto:2008p1683}). 
Both methods give the same 
results within errors, confirming our final estimates. 
For models M1a and M1b we checked the Monte Carlo results 
using integral-equation methods---
the hypernetted chain closure\cite{HansenMcDonald} for model M1a 
and the Rogers-Young closure\cite{RY-84} for 
model M1b: we found very good agreement, indicating that these methods are very
accurate for these soft-sphere models.

\begin{table}
\caption{Compressibility factor $Z(\Phi)$
for the models introduced in Sec.~\ref{sec3},
for the tetramer model (t) of Ref.~\protect\onlinecite{DPP-11}, and 
for polymers (p) in the scaling limit.\protect\cite{Pelissetto:2008p1683}
}
\label{Z-table}
\begin{tabular}{cccccccc}
\hline\hline
$\Phi$ &  p   & M1a    &   M1b     &               M2a   &  M2b  &  M2c   &    t  \\
\hline
0.135& 1.187 & 1.18458(1)& 1.17869(1) & 1.18455(1) & 1.18090(1) & 1.18630(1) &
        1.18993(4) \\
0.27 & 1.393 & 1.38167(1)& 1.36439(1) & 1.38084(2) & 1.36758(1) & 1.38379(2) &
        1.39852(6) \\
0.54 & 1.854 & 1.80067(1)& 1.74840(1) & 1.79770(2) & 1.75010(2) & 1.80119(2) &
        1.8499(1) \\
0.81 & 2.371 & 2.23911(1)& 2.14190(1) & 2.23498(2) & 2.13765(2) & 2.23694(2) &
        \\
1.09 & 2.959 & 2.70461(1)& 2.55534(1) & 2.70088(2) & 2.54008(2) & 2.69982(2) &
        2.9090(1) \\
2.18 & 5.634 & 4.55607(2)& 4.18703(1) & 4.56959(4) & 4.08443(5) & 4.55094(3) &
        5.2660(2) \\
4.36 & 12.23 & 8.29709(2)& 7.47886(3) & 8.36007(5) & 6.9679(1)  & 8.31841(4) &
        10.2056(1) \\
\hline\hline
\end{tabular}
\end{table}

\begin{figure}
\begin{center}
\begin{tabular}{c}
\epsfig{file=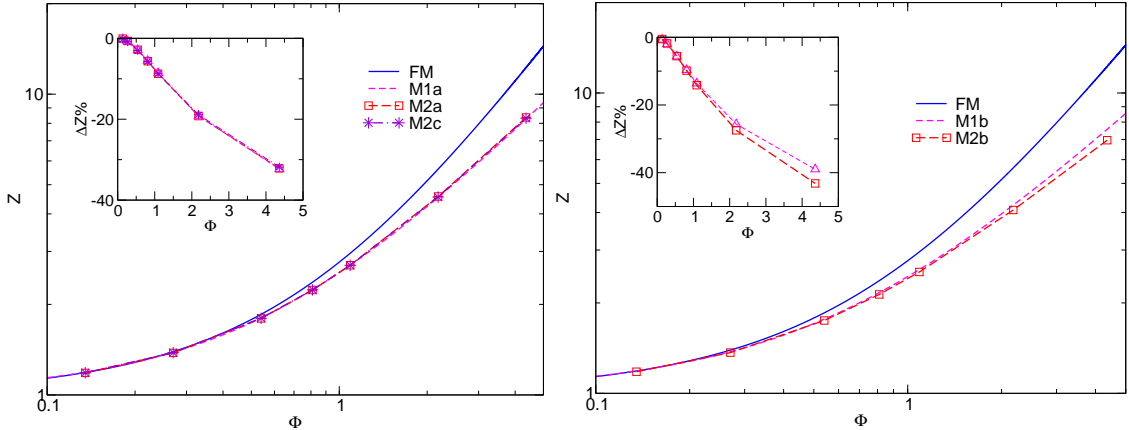,angle=-0,width=15truecm} \hspace{0.5truecm} \\
\end{tabular}
\end{center}
\caption{Compressibility factor $Z$ as a function of $\Phi$.
On the left we report results for models M1a, M2a, M2c, 
on the right we report the results for models M1b, M2b. 
They are compared with the polymer prediction $Z_{FM}$ (full line, FM)
(from Ref.~\protect\onlinecite{Pelissetto:2008p1683}).
In the insets we report the deviations 
$100 (Z/Z_{FM} - 1)$. 
}
\label{EOS-Phi}
\end{figure}

In Table \ref{Z-table} and Fig.~\ref{EOS-Phi} 
we compare the estimates of $Z$ for the different models.
It is evident that considering radii dependent potentials is 
irrelevant for the thermodynamics, as already observed in the discussion of 
$A_3$: models M1a, M2a, and M2c give completely equivalent estimates 
and so do models M1b and M2b. 
Second, the CG model which uses the center of mass as reference
point appears to be more accurate than that using the central monomer.
For $\Phi = 1.09$, which is the expected boundary of applicability of 
single-blob models, models M1a and M2a predict $Z$ with an error of 
9\%, while model M2b underestimates $Z$ by 14\%.

\begin{figure}
\begin{center}
\begin{tabular}{c}
\epsfig{file=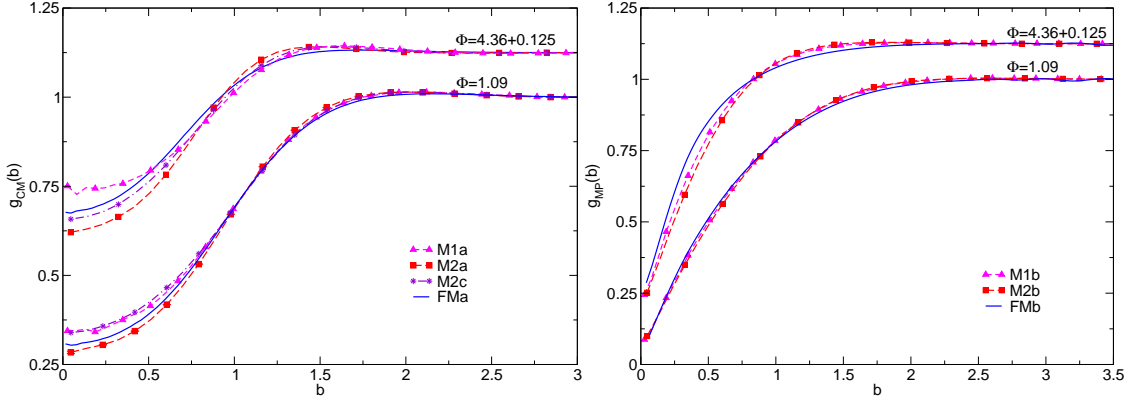,angle=-0,width=15truecm} \hspace{0.5truecm} \\
\end{tabular}
\end{center}
\caption{Intermolecular distribution function for several models at
$\Phi = 1.09$ and $4.36$ (the corresponding function is shifted 
upward for clarity). 
On the left we report results for models M1a, M2a, M2c, 
and the polymer center-of-mass distribution from full-monomer simulations
(FMa);
on the right we report the results for models M1b, M2b, 
and the polymer distribution function associated with the 
polymer midpoint (FMb).
}
\label{gr-Phi}
\end{figure}

It is also interesting to compare the intermolecular distribution functions.
In Fig.~\ref{gr-Phi} we report the
center-of-mass pair distribution function $g_{CM}(b)$ and the 
polymer-midpoint pair distribution function $g_{MP}(b)$.
In the center-of-mass case (left panel) all results are in reasonable
agreement for $b\gtrsim 1$, while larger discrepancies are observed 
for $b \to 0$. For the midpoint distribution all models give similar curves, 
even for small values of $b$. This is due to the logarithmic divergence of 
the potentials, which enforces the condition $g_{MP}(b)\to 0$ for 
$b \to 0$ in all models.

\begin{figure}
\begin{center}
\begin{tabular}{c}
\epsfig{file=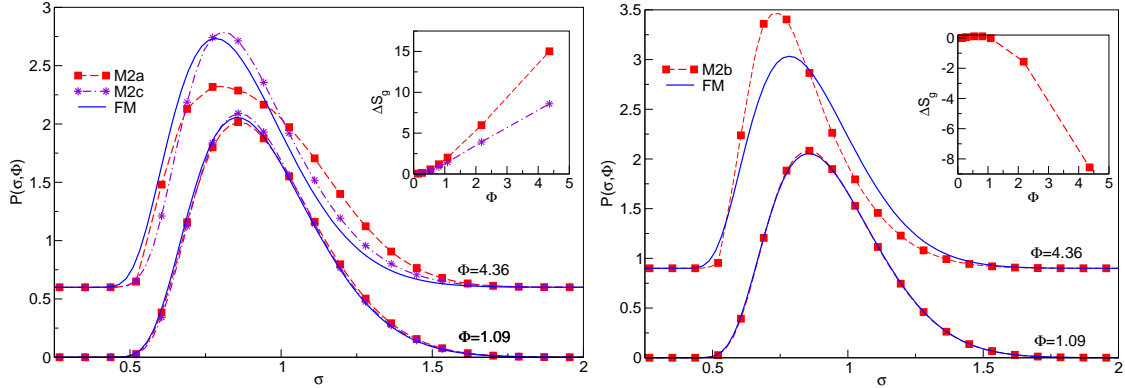,angle=-0,width=15truecm} \hspace{0.5truecm} \\
\end{tabular}
\end{center}
\caption{Distribution $P(\sigma,\Phi)$ of $\sigma = r_g/\hat{R}_g$ for 
$\Phi = 1.09$ and $4.36$ for CG models M2a, M2b, and M2c
and for polymers (FM). In the insets we report the deviations
$\Delta S_g = 100 (S_g/S_{g,FM} - 1)$, where $S_g$ is the ratio 
(\protect\ref{Sg}) for the CG models and $S_{g,FM}$ is the corresponding
quantity for polymers.
}
\label{dist-Rg-Phi}
\end{figure}

Finally, in Fig.~\ref{dist-Rg-Phi} we report the distribution of the radius of 
gyration in the semidilute regime. For $\Phi = 1.09$ both types of 
coarse-graining reproduce correctly the polymer distribution of $r_g$. 
For $\Phi = 4.36$ deviations are significantly larger. However, this 
should not be considered as a problem of the method, but rather as 
consequence of the single-blob model, which is only expected to work
in the dilute regime. To perform a more quantitative comparison 
we consider the ratio 
\begin{eqnarray}
   S_g(\Phi) = \langle r^2_g\rangle_\Phi/\hat{R}_g^2,
\label{Sg}
\end{eqnarray}
where the average is performed at polymer volume fraction $\Phi$. 
For $\Phi = 1.09$ we obtain $S_g(\Phi) = 0.94161(5)$ (M2a), 
0.92300(6) (M2b), 0.93630(6) (M2c), to be compared with the polymer result
$S_g(\Phi) = 0.9238(2)$. In all cases differences are small, although 
model M2b appears to reproduce better the polymer results. 
For larger values of $\Phi$ discrepancies increase, see the insets in 
Fig.~\ref{dist-Rg-Phi}, but this is not surprising for a single-blob model.

\section{Conclusions} \label{sec.4}

In this paper we have performed a detailed study of  single-blob models
\cite{Vettorel:2010p1733} which are characterized by a fluctuating sphere 
radius and by density-independent potentials which are such to reproduce
the radius-of-gyration distribution for an isolated polymer and the 
center-of-mass (or midpoint) polymer distribution function in the limit
of zero polymer density. The results have been compared with those obtained
in simpler single-blob models with fixed blob radius with 
the purpose of understanding if CG models based on compressible blobs 
provide a more accurate description of the thermodynamic behavior 
than fixed-radius
CG models. Since we use single-blob models, this comparison 
can only be made in the dilute regime
$\Phi\lesssim 1$, in which polymer overlaps are rare.

As far as the thermodynamics and the intermolecular structure are concerned,
we find that CG models that use the radii-dependent potentials behave no 
better than those in which polymers are represented by fixed-size spheres.
However, the models with compressible soft spheres allow one to 
reproduce correctly (at least in the dilute regime in which single-blob
models are expected to be predictive)
the density dependence of the  radius of gyration. 
For this quantity, the midpoint representation gives more accurate results.

The fact that models with fluctuating sphere size and models with fixed sphere 
size have the same thermodynamic behavior is probably related to the fact that
the radius of gyration is not the relevant length scale at finite density. 
If one wishes to improve the accuracy of the model, one should take 
into account overlaps which are characterized by a different length scale,
the de Gennes-Pincus correlation length $\xi$, which scales as
$\xi\sim R_g \Phi^{-\gamma}$ with $\gamma=\nu/(3\nu-1) \approx 0.77$
in the semidilute limit.\cite{deGennes,Doi,desCloizeauxJannink,Schaefer}
To describe the
semidilute regime, only multiblob approaches 
appear to be viable coarse-graining methods.
In this respect, on the basis of the present results,
we do not expect the approach of 
Vettorel {\em et al.} \cite{Vettorel:2010p1733} to be more accurate for 
the thermodynamics 
than the more 
straightforward approach of Refs.~\onlinecite{Pierleoni:2007p193,DPP-11}:
considering compressible soft blobs should not provide a model which gives a 
more accurate description of the polymer thermodynamics than those in which
the blob size is fixed. Indeed, both representations equally fail to take 
into account blob overlaps. Of course, a compressible soft-blob model would 
reproduce better the density dependence of some structural properties, like 
the average radius of gyration and the form factor.

One of the difficulties of multiblob approaches is the determination of the 
intra- and inter-molecular 
potentials.\cite{Pierleoni:2007p193,Pelissetto:2009p287,DPP-11}
For models with compressible blobs, a direct numerical determination
appears unfeasible, hence
phenomenological approaches must be used. Ref.~\onlinecite{Vettorel:2010p1733} 
proposed a simple parametrization for the intermolecular blob potential,
Eq.~(\ref{VVBK}). For the single-blob case we find that this parametrization 
is quite accurate:
model M2c is essentially equivalent to model M2a. 

Finally, we have compared the results for two different types of CG models.
As for the thermodynamics, 
CG models for linear polymers that use the center of mass 
as reference point are significantly better than those that use the 
central monomer. Indeed, the estimates of the third virial coefficient and of 
the compressibility factor corresponding to models M1a and M2a are 
significantly closer to the full-monomer estimates than those 
obtained by using models M1b and M2b. 

\section*{Acknowledgements}

C.P. is supported by the Italian Institute of Technology (IIT) under the
SEED project grant number 259 SIMBEDD – Advanced Computational Methods for
Biophysics, Drug Design and Energy Research.

\appendix 
\section{Explicit expression for the third virial coefficient}

In this Appendix we wish to derive an expression for the third virial
coefficient which explicitly depends on the three-body potential of 
mean force defined in Eq.~(\ref{def-V3}). 
We consider a generic system of molecules with internal degrees of freedom 
and on each molecule we select a reference point $S$. Then, we indicate 
with $\langle \cdot \rangle_{\bf r}$ the average over all internal
degrees of freedom of the molecule such that the position of point $S$ is 
${\bf r}$. The formalism can be applied both to polymers --- in this case
the average is over all polymer conformations and $S$ can be taken as 
the polymer center of mass or the central monomer --- and to the 
single-blob CG model --- in this case $S$ is the position of 
the sphere and the average is over all values of its radius.  Analogously, we define 
$\langle \cdot \rangle_{{\bf r}_1,{\bf r}_2}$ as the average over all additional 
degrees of freedom of two isolated molecules such that 
$S_1$ is in ${\bf r}_1$ and $S_2$ is in ${\bf r}_2$, and similar 
expressions involving three molecules. Finally, we define the zero-density
$S$-related pair distribution function
\begin{equation}
g_{S}(r) = \langle e^{-\beta U_{12}}  \rangle_{{\bf 0},{\bf r}},
\end{equation}
where $U_{12}$ is the intermolecular potential energy, and the 
corresponding correlation function
\begin{equation}
h_S(r) = g_{S}(r) - 1 = \langle e^{-\beta U_{12}} - 1 \rangle_{{\bf 0},{\bf r}}
       = \langle f_{12} \rangle_{{\bf 0},{\bf r}},
\end{equation}
where $f_{12}$ is the usual Mayer function.

The third virial coefficient $B_3$ for the model can be written as 
\cite{Caracciolo:2006p587}
\begin{equation}
B_3 = - {1\over 3} I_3 - T_1 ,
\end{equation}
where 
\begin{eqnarray}
I_3 &=& \int d^3{\bf r}_{12} d^3{\bf r}_{13}
   \langle f_{12} f_{13} f_{23} \rangle_{{\bf 0},{\bf r}_{12},{\bf r}_{13}},
\nonumber \\
T_1 &=& \int d^3{\bf r}_{12} d^3{\bf r}_{13} \left(
   \langle f_{12} f_{13}  \rangle_{{\bf 0},{\bf r}_{12},{\bf r}_{13}} - 
   \langle f_{12} \rangle_{{\bf 0},{\bf r}_{12}} 
   \langle f_{13} \rangle_{{\bf 0},{\bf r}_{13}} \right).
\end{eqnarray}
Using definition (\ref{def-V3}) we can rewrite
\begin{eqnarray}
&& \langle f_{12} f_{13} f_{23} \rangle_{{\bf 0},{\bf r}_{12},{\bf r}_{13}} 
   = \left(e^{-\beta V_3({\bf r}_{12},{\bf r}_{13},{\bf r}_{23})} - 1\right)
     g_S(r_{12}) g_S(r_{13}) g_S(r_{23}) \nonumber \\
&& \qquad + 
     h_S(r_{12}) h_S(r_{13}) h_S(r_{23}) \nonumber \\
&& \qquad - 
    \left(
   \langle f_{12} f_{13}  \rangle_{{\bf 0},{\bf r}_{12},{\bf r}_{13}} -
     h_S(r_{12}) h_S(r_{13})  + 
   \hbox{2 permutations}\right),
\end{eqnarray}
where ${\bf r}_{23}= {\bf r}_{13} - {\bf r}_{23}$ and the two permutations
correspond to replacing once $13$ with $23$ and the second time $12$ with $23$.
Using this expression we end up with 
\begin{eqnarray}
B_3 &=& - {1\over 3} \int d^3{\bf r}_{12} d^3{\bf r}_{13} 
\left(e^{-\beta V_3({\bf r}_{12},{\bf r}_{13},{\bf r}_{23})} - 1\right) 
   g_S(r_{12}) g_S(r_{13}) g_S(r_{23})
\nonumber \\
    && - {1\over 3} \int d^3{\bf r}_{12} d^3{\bf r}_{13}
      h_S(r_{12}) h_S(r_{13}) h_S(r_{23}).
\label{B3-V3}
\end{eqnarray}
It is interesting to observe that this expression is identical to the one 
which applies to a system of monoatomic molecules (without additional degrees
of freedom) interacting by means of a two-body and of a three-body potential.

Using Eq.~(\ref{B3-V3}) we can compute the difference 
$B_{3,\rm pol} - B_{3,CG}$ of the third virial coefficient for polymers 
and for the CG model. Since the $S$-related pair distribution
function is, by definition, the same in the two models, the difference 
is given by Eq.~(\ref{B3diff}) reported in the main text with 
$g(r) = g_S(r)$.

\providecommand{\newblock}{}


\begin{thebibliography}{100}
\bibitem{footnotevir}
In the low-density limit the (osmotic) pressure $\Pi$ of a polymer solution 
has an expansion of the form $\Pi/k_B T = c + B_2 c^2$, where 
$c = L/V$ is the number of polymers per unit volume. 
The combination $A_2 = B_2 \hat{R}_g^3$ is universal.
\end{thebibliography}
\end{document}